\documentstyle[e-e-ijmpa,twoside,epsfig]{article}

\makeatletter
\def\@cite#1#2{{[{#1}]\if@tempswa\typeout
{IJCGA warning: optional citation argument
ignored: `#2'} \fi}}
\newcount\@tempcntc
\def\@citex[#1]#2{\if@filesw\immediate\write\@auxout{\string\citation{#2}}\fi
  \@tempcnta\z@\@tempcntb\m@ne\def\@citea{}\@cite{\@for\@citeb:=#2\do
    {\@ifundefined
       {b@\@citeb}{\@citeo\@tempcntb\m@ne\@citea\def\@citea{,}{\bf ?}\@warning
       {Citation `\@citeb' on page \thepage \space undefined}}%
    {\setbox\z@\hbox{\global\@tempcntc0\csname b@\@citeb\endcsname\relax}%
     \ifnum\@tempcntc=\z@ \@citeo\@tempcntb\m@ne
       \@citea\def\@citea{,}\hbox{\csname b@\@citeb\endcsname}%
     \else
      \advance\@tempcntb\@ne
      \ifnum\@tempcntb=\@tempcntc
      \else\advance\@tempcntb\m@ne\@citeo
      \@tempcnta\@tempcntc\@tempcntb\@tempcntc\fi\fi}}\@citeo}{#1}}
\def\@citeo{\ifnum\@tempcnta>\@tempcntb\else\@citea\def\@citea{,}%
  \ifnum\@tempcnta=\@tempcntb\the\@tempcnta\else
   {\advance\@tempcnta\@ne\ifnum\@tempcnta=\@tempcntb \else \def\@citea{--}\fi
    \advance\@tempcnta\m@ne\the\@tempcnta\@citea\the\@tempcntb}\fi\fi}
\makeatother

\renewcommand{\thefootnote}{\fnsymbol{footnote}}	%USE SYMBOLIC FOOTNOTE

\def\sighmmbar{\overline\sigma_{\hmm}}
\def\gamhmm{\Gamma_{\hmm}^{\rm tot}}

\def\thetaw{\theta_W}
\def\gamz{\Gamma_Z}
\def\xw{x_W}
\def\yw{y_W}

\def\hmp{\h^{-,+}}

\def\vev#1{\langle #1 \rangle}

\def\slep{\wt \ell}

\def\slepl{\wt \ell_L}

\def\h{H}
\def\a{A}

\def\eg{{\it e.g.}}

\def\dmm{\Delta^{--}}

\def\hzero{\h^0}

\def\hmm{H^{--}}
\def\mhmm{m_{\hmm}}
\def\hpp{H^{++}}

\def\hppmm{H^{++,--}}

\def\stop{\wt t}

\def\mstop{m_{\stop}}

\def\slep{\wt \ell}

\def\slepl{\wt \ell_L}

\def\eg{{\it e.g.}}

\def\stop{\wt t}

\def\mstop{m_{\stop}}

\def\slep{\wt \ell}

\def\slepl{\wt \ell_L}

\def\hsm{h_{\rm SM}}

\def\hl{h^0}
\def\hh{H^0}
\def\ha{A^0}
\def\hp{H^+}
\def\hm{H^-}
\def\hpm{H^{+,-}}

\def\mhh{m_{\hh}}
\def\mha{m_{\ha}}

\def\mhpm{m_{\hpm}}
\def\tanb{\tan\beta}

\def\mz{m_Z}
\def\mw{m_W}
\def\mgut{M_U}

\def\wp{W^+}
\def\wm{W^-}
\def\wpm{W^{+,-}}
\def\wmp{W^{-,+}}

\def\h{H}

\def\wt{\widetilde}

\def\cpmone{\wt \chi^{\pm}_1}

\def\mcpmone{m_{\cpmone}}

\def\emem{e^-e^-}
\def\mummum{\mu^-\mu^-}
\def\dmm{\Delta^{--}}

\def\hhmm{h^{\dmm}}

\def\hzero{\h^0}

\def\MPL #1 #2 #3 {Mod.~Phys.~Lett.~{\bf#1},\  #2 (#3)}
\def\NPB #1 #2 #3 {Nucl.~Phys.~{\bf#1},\  #2 (#3)}
\def\PLB #1 #2 #3 {Phys.~Lett.~{\bf#1},\  #2 (#3)}
\def\PR #1 #2 #3 {Phys.~Rep.~{\bf#1},\ #2 (#3)}
\def\PRD #1 #2 #3 {Phys.~Rev.~{\bf#1},\  #2 (#3)}
\def\PRL #1 #2 #3 {Phys.~Rev.~Lett.~{\bf#1},\  #2 (#3)}
\def\RMP #1 #2 #3 {Rev.~Mod.~Phys.~{\bf#1},\  #2 (#3)}
\def\ZP #1 #2 #3 {Z.~Phys.~{\bf#1},\  #2 (#3)}
\def\IJMP #1 #2 #3 {Int.~J.~Mod.~Phys.~{\bf#1},\  #2 (#3)}
\def\IBID #1 #2 #3 {{\bf#1},\  #2 (#3)}

\def\call{{\cal L}}

\def\tauptaum{\tau^+\tau^-}

\def\br{BF}
\def\tauptaum{\tau^+\tau^-}

\def\gam{\gamma}
\def\sigrts{\sigma_{\tiny\rts}^{}}

\def\anti{\overline}
\def\epem{e^+e^-}

\def\mupmum{\mu^+\mu^-}
\def\lplm{\ell^+\ell^-}
\def\lmlp{\ell^-\ell^+}
\def\lmlm{\ell^-\ell^-}

\def\rts{\sqrt s}
\def\ie{{\it i.e.}}
\def\eg{{\it e.g.}}

\def\anti{\overline}
\def\wp{W^+}
\def\wm{W^-}
\def\mw{m_W}
\def\mz{m_Z}
\def\h{H}

\def\a{A}

\def\hsm{h_{SM}}

\def\tanb{\tan\beta}
\def\hl{h^0}

\def\ha{A^0}
\def\mha{m_{\ha}}

\def\hh{H^0}
\def\mhh{m_{\hh}}

\def\fbi{~{\rm fb}^{-1}}

\def\mev{~{\rm MeV}}
\def\gev{~{\rm GeV}}
\def\tev{~{\rm TeV}}
\def\stop{\widetilde t}
\def\mstop{m_{\stop}}

\def\dmm{\Delta^{--}}

\def\hzero{\h^0}

\newcommand{\nc}{\newcommand}
\nc{\beq}{\begin{equation}}   \nc{\eeq}{\end{equation}}
\nc{\bea}{\begin{eqnarray}}   \nc{\eea}{\end{eqnarray}}
\nc{\baa}{\begin{array}}      \nc{\eaa}{\end{array}}
\nc{\bit}{\begin{itemize}}    \nc{\eit}{\end{itemize}}
\nc{\ben}{\begin{enumerate}}  \nc{\een}{\end{enumerate}}
\nc{\bce}{\begin{center}}     \nc{\ece}{\end{center}}

\def\overlay#1#2{\ifmmode \setbox 0=\hbox {$#1$}\setbox 1=\hbox to\wd 0{\hss
$#2$\hss }\else \setbox 0=\hbox {#1}\setbox 1=\hbox to\wd 0{\hss #2\hss }\fi
#1\hskip -\wd 0\box 1}
\def\case#1/#2{{\textstyle{#1\over#2}}}
\def\9{\phantom 0}      %%% for lining up numbers in columns
\renewcommand\linebreak{\unskip\break} %% breaks line & still justifies
\newcommand{\alt}{\mathrel{\raisebox{-.6ex}{$\stackrel{\textstyle<}{\sim}$}}}
\newcommand{\agt}{\mathrel{\raisebox{-.6ex}{$\stackrel{\textstyle>}{\sim}$}}}
\def\lsim{\alt}
\def\gsim{\agt}
\begin{document}

\font\fortssbx=cmssbx10 scaled \magstep2
\hbox to \hsize{
$\vcenter{
\hbox{\fortssbx University of California - Davis}
}$
\hfill
$\vcenter{
\hbox{\bf UCD-98-6} 
\hbox{\bf hep-ph/9803222}
\hbox{March, 1998}
}$
}
\medskip

\normalsize\textlineskip
\pagestyle{empty}

\title{Probing Exotic Higgs Sectors in {\boldmath $\lmlm$}
Collisions
\footnote{To appear in {\it $e^-e^-$ 1995: Proceedings of
the Electron-Electron Linear Collider Workshop}, Santa Cruz, California,
September, 1997, edited by C. Heusch, to be published in
Int.~J.~Mod.~Phys.~{\bf A}.}
}

\author{John F. Gunion
\footnote{Work supported in part by the Department of Energy, 
Contract  DE-FG03-91ER40674, and by the Davis Institute for High Energy
Physics.   }
}
\address{Davis Institute for High Energy Physics \\
Department of Physics,
University of California at Davis, Davis CA 95616}

\maketitle\abstracts{I review extended
Higgs sectors and constraints thereon arising
from $\rho=1$, gauge-coupling unification and $b\to s\gam$.
The couplings and decays of the Higgs boson eigenstates are outlined
for triplet representations.
Direct experimental probes of exotic Higgs bosons are reviewed with
a focus on the important role that would be played by an $e^-e^-$
or $\mu^-\mu^-$ collider.}

\setcounter{footnote}{0}
\renewcommand{\thefootnote}{\alph{footnote}}

\vspace*{1pt}\textlineskip	

\section{Introduction}

If the origin of electroweak symmetry breaking lies in the existence of
a Higgs sector, understanding the full nature of this sector will be
one of the primary goals of future experimental programs.
Although the Higgs sector could be quite simple, it is also possible
that it will yield some real surprises. The purpose of this brief
review is to assess the attractiveness of some exotic possibilities
and the role of a $\lmlm$ collider in exploring them.
It will be convenient to divide the
possible Higgs representations into:
(a) singlets; (b) doublets; (c) triplets; (d) higher representations.
The most important current aesthetic and experimental constraints
on the Higgs representations, include: (i)
naturality of $\rho=1$; (ii) gauge coupling unification; and
(iii) the branching ratio
for $b\to s\gam$. After reviewing these items as they
affect various representations, I will turn to
future experimental probes in high energy collisions,
focusing on the importance of $\lmlm$ collisions.

It is useful to list a selection of possible representations.
For the moment, I restrict the discussion to the Standard Model (SM)
SU(3)$\otimes$SU(2)$_L$$\otimes$U(1) gauge group, with $Q=I_3+Y/2$.\footnote{I
will employ the notation $I$ for weak SU(2)$_L$ isospin.}~ 
In the case of exotic representations, I will emphasize
those which would lead to a doubly-charged Higgs boson eigenstate.
A doubly-charged Higgs boson 
would constitute an incontrovertible signature for an
exotic representation with non-standard U(1) hypercharge
and/or SU(2)$_L$ weak isospin $I\geq 1$ and would very likely be of
particular interest for $\lmlm$ collisions. 
For simplicity, the Higgs sector will be assumed to
be CP-conserving, with CP-even
neutral states denoted by $h$ or $\h$ and CP-odd states by $a$ or $\a$.
The notation will be S$_Y$ for
an $I=0$ singlet, D$_Y$ for an $I=1/2$ doublet, and T$_Y$ for an $I=1$ triplet
representation, respectively, where
$Y$ denotes the absolute value of the hypercharge. All representations
will be complex unless otherwise stated. 
%\clearpage
\bit 
\item
Doublet Models:
\eit
\begin{description}
\item[--] 
The 1D$_1$ minimal Standard Model (MSM) Higgs sector with a single
Higgs eigenstate, the $\hsm$.
\item[--] 
A 2D$_1$ model, yielding the CP-even $\hl$ and $\hh$, a CP-odd $\ha$, and
a $\hpm$ pair. In this review, only type-II two-doublet
models (\ie\ ones in which one doublet, $H_u$, 
gives mass to up-type quarks and the
other, $H_d$, gives mass to down-type quarks and leptons)
are considered. An important parameter of the model is the ratio
$\tanb=\vev{H_u}/\vev{H_d}$ of the neutral Higgs field component vev's.
If $\mha$ is large, it is natural, but not required, that
the two-doublet model exhibit decoupling, according to which the light $\hl$
becomes SM-like while the $\hh$, $\ha$ and $\hpm$ decouple from electroweak
symmetry breaking.
\item[--] 
The 2D$_1$ minimal supersymmetric model (MSSM) \cite{ghseries}. 
Supersymmetry imposes
restrictions on the Higgs sector which guarantee that the $\hl$
becomes SM-like when $\mha\sim\mhh\sim\mhpm$ becomes large.
\item[--] 
A 2D$_1$+1S$_0$ model, yielding states $\h_{1,2,3}$, $\a_{1,2}$, and $\hpm$.
\item[--] A 1D$_1$+1D$_{Y>1}$ model. The physical states
depend on $|Y|$; \eg\ for $|Y|=3$ we have 
$\hsm$, $\hpm$,   and $\hppmm$.
\item[--] A 1D$_1$+1S$_4$ model, yielding  $\hsm$ and $\hppmm$.
\end{description}

\bit
\item
Triplet Models plus at least 1D$_1$ for fermion masses:
\eit

\begin{description}
\item[--] 
1D$_1$+1T$_Y$ models; the physical states depend on the $Y$ of the triplet.
For example, a
$Y=0$ (real representation) yields CP-even neutral states $\h_{1,2}$ and a 
$\hpm$ pair.
A $Y=2$ (complex representation) yields $\h_{1,2}$, $\a$, $\hpm$, $\hppmm$.
\item[--] A 1D$_1$+1T$_2$+1T$_0$(real) Higgs sector yields
$\h_{1,2,3}$, $\a$, $\hpm_{1,2}$, $\hppmm$. 
\item[--] A 1D$_1$+2T$_2$ model yields $\h_{1,2,3}$, $\a_{1,2}$, 
$\hpm_{1,2}$, and $\hppmm_{1,2}$.
\end{description}

\bit
\item Higher Representations:
\eit

\begin{description}
\item[--] 
The most interesting possibility is 
a 1D$_1$+1($I=3,|Y|=4$) Higgs sector, which yields  
$\h_{1,2}$, $\a$, $\hpm_{1,2}$, $\hppmm$,
$\h^{+++,---}$, $\h^{++++,----}$, $\h^{+++++,-----}$ 
\end{description}

The 1D$_1$+1T$_2$+1T$_0$(real) Higgs sector
\cite{georgiplus,georgichiv,mscmgtri,trihtree,trihloop,bamkun} with
equal vev's for the T$_2$ and T$_0$ triplet representation
neutral members provides
an especially useful benchmark. As summarized in Refs.~\cite{trihtree,hhg},
the Higgs eigenstates can be separated into a weak isospin
five-plet ($H_5^{++,--}$,
$H_5^{+,-}$, $H_5^0$), a triplet ($H_3^{+,-}$, $H_3^0$) and two
singlets ($H_1^0$, $H_1^{0\,\prime}$). Of these, only the
$H_3$'s and the $H_1^0$ have overlap with the doublet Higgs fields
and, therefore, have fermionic couplings. An important parameter
in this model is $\tan\theta_H$ which is proportional to the ratio
of the (common) vev of the neutral triplet members 
to the vev of the neutral doublet member.

\section{Aesthetic/Experimental Constraints}

\bit
\item $\rho=1$ at tree-level:
\eit

The first major consideration is the naturalness of
$\rho=\mw^2/[\mz^2\cos^2\theta_W]=1$. The starting point
is the well-known result:
\beq
\rho={\sum_{I,Y}\left[4I(I+1)-Y^2\right]|v_{I,Y}|^2 c_{I,Y}\over
\sum_{I,Y}2Y^2|v_{I,Y}|^2}\,,
\label{rhodef}
\eeq
where $I$ specifies the SU(2)$_L$ representation,
$v_{I,Y}$ is the vev of the neutral member of the representation {\it if any},
and $c_{I,Y}=1 (1/2)$ for complex (real $Y=0$) representations, respectively.
The two lowest single representation solutions to $\rho=1$ are
$I=1/2,|Y|=1$ and $I=3,|Y|=4$. The minimal SM employs a single $I=1/2,|Y|=1$
doublet representation. As delineated earlier, a 1D$_1$+1$(I=3,|Y|=4)$
Higgs sector would have an extensive array of eigenstates.

If a neutral member of a SU(2)$_L$ triplet representation 
acquires a non-zero vev,
$\rho=1$ is never automatic. Even in the 1D$_1$+1T$_2$+1T$_0$(real)
model with equal neutral vev's for the triplets,
which has $\rho=1$ at tree-level, 1-loop corrections to $\rho$ are
infinite \cite{mscmgtri,trihloop}.  
This means that $\rho$ is a renormalizable quantity,
the value of which must be inserted into the theory
as an additional experimental input (just like $\alpha$,
$G_F$, $\ldots$ in the SM) \cite{trihloop}. 
The precision electroweak data has been analyzed in this context 
\cite{hollik} and it is found
that there is no problem fitting all data, but this is hardly surprising
given the loss of a (generally very constraining) prediction for $\rho$.
In order to maintain predictability for $\rho$, 
it is tempting to favor models in which
$v_L=0$ for any L-triplet neutral member.
(Note that $v_R\neq 0$ in L-R symmetric models is possible
without affecting $\rho$, and is necessary
for $m_{W_R}\gg m_{W_L}$. Of course, $m_{W_R}$ is then a free renormalizable
parameter.)

Among the exotic representations listed earlier, those
which do not contain a neutral member are interesting in that $\rho=1$
remains natural if such a representation is included in the Higgs sector.

\bit
\item Gauge coupling unification:
\eit

\def\ndi{N_{D_1}}
\def\ndiii{N_{D_3}}
\def\nto{N_{T_0}}
\def\ntii{N_{T_2}}
\def\nsii{N_{S_2}}
\def\nsiv{N_{S_4}}
\def\n34{N_{34}}
The requirement that the gauge couplings unify with a desert between
the TeV energy scale and the GUT unification scale $\mgut$
is widely regarded as being highly desirable. 
The resulting constraints on the Higgs representations that can be present
are easily determined. At 1-loop, gauge 
coupling evolution depends only on the couplings themselves and
the Higgs and other particle 
representations present at any given energy scale. Let us denote
the number of $|Y|=1$ doublets by $\ndi$, the number of $|Y|=2$ triplets 
by $\ntii$, and so forth; $\n34$ denotes the number of $(I=3,|Y|=4)$
representations. I will not consider
$|Y|\geq 6$ singlets, $|Y|\geq 5$ doublets, or $|Y|\geq 4$ triplets.
The 1-loop evolution coefficients are then as follows (using $N_g$
to denote the number of complete generations).

\medskip
\noindent\underline{Non-SUSY, MSM:}
\bea
&b_1={4\over 3}N_g+{1\over 5}(\nsii+4\nsiv)+
{1\over 10}(\ndi+9\ndiii)+{3\over 5}\ntii+{28\over 5}\n34&\nonumber\\
&b_2={4\over 3}N_g+{1\over 6}(\ndi+\ndiii)+{2\over
3}(\nto+\ntii)+{28\over27}\n34-{22\over 3}&\nonumber \\
&b_3={4\over 3}N_g-11\,.&
\label{nonsusybs}
\eea
\noindent\underline{SUSY, MSSM sparticle content only:}
\bea
&b_1=2N_g+{3\over 5}(\nsii+4\nsiv)+
{3\over 10}(\ndi+9\ndiii)+{9\over 5}\ntii+{84\over 5}\n34&\nonumber\\
&b_2=2N_g+{1\over 2}(\ndi+\ndiii)+2(\nto+\ntii)+{28\over9}\n34 -6&\nonumber\\
&b_3=2N_g-9&\,.
\label{susybs}
\eea
Note that in the above equations there is
no influence from $Y=0$ singlets, but $Y\neq 0$ singlets affect $b_1$.

The requirement of unification (assuming ``standard'' SU(5) normalization of 
the U(1) coupling constant and a desert between $\mz$
and $\mgut$) can be written (at 1-loop) in the form
\beq
\alpha_s(\mz)=\alpha_{QED}(\mz) {5(b_1-b_2)\over
\sin^2\theta_W(5b_1+3b_2-8b_3)-3(b_2-b_3)}
\label{unificond}
\eeq
Using $\alpha_s(\mz)=0.118$ and $\sin^2\theta_W=.2315$,
and the results of Eqs.~(\ref{nonsusybs}) and
(\ref{susybs}), the constraint of Eq.~(\ref{unificond}) reduces to
\bea
&\phantom{S}{\rm SM}:~~1\simeq -0.09\nsii-0.36\nsiv+0.13\ndi-0.22\ndiii&
\nonumber\\
&\phantom{S{\rm SM}:~~1\simeq} +0.71\nto+0.44\ntii-1.39\n34&
\label{smconstraints}\\
&{\rm SUSY}:~~1\simeq -0.33\nsii-1.31\nsiv+0.49\ndi-0.82\ndiii&\nonumber\\
&\phantom{{\rm SUSY}:~~1\simeq}+2.61\nto+ 1.63\ntii-5.11\n34 &
\label{susyconstraints}
\eea
in the non-SUSY and MSSM cases, respectively.  
The exact coefficients in the equations above are sensitive 
to the precise $\alpha_s$ and $\sin^2\theta_W$ choices as well as
to whether all the Higgs bosons have mass near $\mz$, as assumed,
or nearer to 1 TeV.
Two-loop corrections also lead to small changes in the unification conditions.
Thus, the following discussion of `solutions' should be regarded as being
a somewhat rough, but indicative, guide to the possibilities.

In assessing the constraints of Eqs.~(\ref{smconstraints}) and
(\ref{susyconstraints}), it should be kept in mind that
$\ndi\geq 1$ ($\geq 2$) is required in the SM (MSSM)
for Dirac fermion masses, and that 
in the SUSY context, all $|Y|\neq 0$ (complex) multiplets must come 
in $Y=\pm |Y|$ pairs (to cancel anomalies).
Although there are many solutions to the equations, essentially all
but the two-doublet MSSM solution 
are excluded if one requires a large value for $\mgut$,\footnote{This
extends the result of \cite{amaldi} where
small $\mgut$ was found for multi-doublet SM and $\ndi\geq 4$ 
MSSM solutions.}~ 
as would be needed to ensure proton stability if the SU(3), SU(2)
and U(1) groups merge to form a single group [such as SU(5)] 
containing gauge bosons that mediate proton decay.
In what follows, a solution is defined as being any choice for the Higgs
representations that yields $0.1\leq \alpha_s\leq 0.13$. For simplicity,
only values with $\ndi\leq 16$ and all other $N$'s $\leq 8$ have
been surveyed.
\begin{enumerate}
\item 
In the SM, the highest $\mgut$ value is attained for the
$\ndi=2$, $\nto=1$ solution,\footnote{$N$'s not stated
are all 0.}~ 
yielding $\alpha_s=0.115,\mgut\sim 1.6\times 10^{14}\gev$. There
are some 35 additional solutions with $\mgut\geq 10^{13}\gev$,
the next highest $\mgut$ solutions being: $\alpha_s=0.104,\mgut\sim 1.2\times
10^{14}\gev$ for $\ndi=1,\nto=1$ and $\alpha_s=0.120,\mgut\sim 10^{14}\gev$
for $\ndi=3,\nto=1$, which are obviously closely related to the best
solution. Next on the list are $\ndi=2,\nto=1,\nsii=1$, yielding
$\alpha_s=0.108,\mgut\sim 7\times 10^{13}\gev$ and $\ndi=8$, yielding
$\alpha_s=0.124,\mgut\sim 5\times 10^{13}\gev$. The $\ntii\neq0$
solution with largest $\mgut$ is $\ndi=3,\ntii=1$, yielding
$\alpha_s=0.105,\mgut\sim 2\times 10^{13}\gev$.
All $\ndiii\neq0$ solutions have $\mgut< 10^{13}\gev$.
The $\nsiv\neq0$ solutions with highest $\mgut$ are $\ndi=1,\nsiv=1$
and $\nsii=1$ or $2$, yielding $\alpha_s=0.129,\mgut\sim 3\times 10^{13}\gev$
or $\alpha_s=0.120,\mgut\sim 1.5\times 10^{13}\gev$, respectively.
The D$_1$ and $(I=3,|Y|=4)$ representations are especially interesting
since any combination of such representations
would yield $\rho=1$ naturally even when their neutral members
have non-zero vev's. However, the
$\ndi=18,\n34=1$ solution yields $\mgut\sim 2.8\times 10^9\gev$; higher
values of $\n34$ yield still smaller $\mgut$ values.
\item 
In SUSY, the $\ndi=2$ solution to Eq.~(\ref{susyconstraints}) 
with $\alpha_s=0.115,\mgut\gsim 2 \times 10^{16}\gev$ is far and away the best.
Next highest $\mgut$ is achieved for the $\ndi=2,\nto=1,$\footnote{Since
the T$_0$ representation is real, it is automatically anomaly-free;
thus any value of $\nto$ is allowed.}~$\nsii=8$
and $\ndi=6,\nsii=6$ solutions, both of which yield
$\alpha_s=0.115,\mgut\sim 3\times 10^{13}\gev$.
Any $\ndiii\geq 2$ or $\nsiv\geq 2$ solution has much smaller $\mgut$.
Solutions with $(I=3,|Y|=4)$ representations inevitably lead to very
small $\mgut$.
For example, the $\ndi=22,\n34=2$ solution yields $\mgut\sim 
5\times 10^5\gev$.
\item 
The L-R symmetric extension of the SM and the supersymmetric
L-R symmetric model both  require intermediate
scale matter for full coupling unification \cite{lrunification}. 
Many new possible solutions
also emerge in the basic SM or SUSY contexts 
by including intermediate scale matter;
in particular, solutions containing exotic
Higgs representations that yield a large value of $\mgut$
become possible. The intermediate-scale-matter solutions
will not be pursued here.
\end{enumerate}

\bit
\item $\rho$ at one-loop:
\eit

In models with $\rho=1$ at tree level and for which $\rho$ is finitely
calculable, parameters must still be 
chosen so that Higgs-loop corrections to $\rho$
are small. Too large a mass separation between
neutral and charged Higgs bosons with $W,Z$ couplings 
will lead to a large $\delta\rho$ at 1-loop. (See \cite{hhg}
for a brief discussion.) A large $\delta\rho$
is automatically avoided in supersymmetric models throughout all
of Higgs parameter space by virtue of the fact that
heavier Higgs bosons decouple from the $W,Z$ sector. For example,
in the minimal two-doublet-Higgs-sector MSSM model, when
$\mha\sim\mhh\sim\mhpm$ is large the dangerous $\wp\to \hp\hl$
and $Z\to\hh\hl$ couplings [proportional to the famous $\cos(\beta-\alpha)$
factor] are highly suppressed \cite{hhg}.

\bit
\item $b\to s\gam$:
\eit

Any Higgs sector containing more than one $|Y|=1$ doublet will
have one or more singly-charged Higgs bosons with fermionic couplings.
Any such $\hp$ will enter into a 1-loop $\hp\anti t$ 
contribution to the $b\to s\gam$
transition that must be added at the amplitude level to the SM $\wp\anti t$
loop.  In the case of a type-II two-doublet model, the $\hp\anti t$
loop adds constructively to the $\wp\anti t$ SM loop. Since the SM $\wp \anti
t$ loop alone leads to a $b\to s\gam$ branching ratio that exceeds the measured
value, the 95\% confidence level limits
on $\mhpm$ are large \cite{joanne}, roughly $\mhpm\gsim 300\gev$
for $\tanb>1$,  unless there are additional
1-loop graphs that can cancel the extra $\hp \anti t$ loop contribution.
The best-known example of such a cancellation
arises in supersymmetric models. There,
a stop-chargino graph can cancel an excessive top-charged-Higgs
graph if $\mstop$ and $\mcpmone$ are small enough. Of course,
charged Higgs that do not couple to quarks (of
which there are many in the models listed earlier) are no problem.

\section{Couplings and Decays}

The phenomenology of exotic Higgs representations 
is a very complex topic and very model-dependent.  
In order to avoid too lengthy a discussion, I will focus 
on models containing triplet representations in addition
to one or more doublet Higgs representations. 
A convenient summary of Higgs triplet phenomenology 
appears in \cite{hhg}; more details can be found, for example,
in \cite{vegadicus,trihtree,lr89,hanemem,jfgemem,rizzoemem,huituetal,cheungetal,godboleetal}.
I begin with a few preliminary reminders of well-known facts.

\begin{itemize}

\item
There is never a $\gamma \wpm\hmp$ vertex. There is generally
a non-zero tree-level $Z\wpm\hmp$ vertex 
if $v_L\neq 0$ (for a $|Y|\neq 0$ representation with non-zero
SU(2)$_L$ isospin), even
if $\rho$ is tuned to $\simeq 1$ at tree-level. 
However, if we require $v_L=0$ in order to maintain naturality of
$\rho=1$, all $Z\wpm\hmp$ vertices will be zero at tree-level.

\item
Charged and neutral (non-singlet) Higgs bosons, triplet members or otherwise,
have diagonal pair 
couplings to $\gam\propto Q$ and $Z\propto I_3-\xw Q$.

\item
$W_LW_L$ couplings to Higgs bosons of an L-triplet
representation are proportional to the vev $v_L$ of 
the neutral member of the triplet. This means that
$\rho=1$ is only natural
if there are no $W_LW_L$ couplings at tree-level. 
In a L-R symmetric model,
$W_RW_R$ couplings to R-triplet members are $\propto v_R$, 
and are generally large.

\item
This same rule applies to multi-triplet models.
For the $H_T$'s of such a model, $H_TVV$ couplings are proportional to an
appropriate triplet vev (or zero if $H_T$ is CP-odd in nature).
As an example, in the 1D$_1$+1T$_2$+1T$_0$(real) Higgs sector
with $\rho=1$ at tree-level 
(as summarized in Ref.~\cite{hhg} and the Introduction) 
all such couplings are proportional to $\sin\theta_H$.

\item
For triplet $H_T$'s, most $H_TH_T^{(\prime)}V$ couplings are non-zero even
if the triplet vev is zero. Again the 1D$_1$+1T$_2$+1T$_0$(real) Higgs sector
provides a useful example. Of particular interest are couplings
of the type $H_T^+H_T^{--}\to \wm$ which can allow for $H_T^+H_T^{--}$
production via $\wm$ exchange and $H_T^{--}\to H_T^-\wm$ decays
(and charge conjugate versions thereof). These couplings are
sizeable, independent of the magnitude of the neutral triplet vev.

\item
Non-zero trilinear couplings of three $H_T$'s require a triplet vev.

\item
$H_DH_TH_T$
couplings involving one doublet $H_D$ and two $H_T$'s are non-zero
if the doublet vev is non-zero, even if the triplet vev is zero.

\item When quantum-number allowed, $H_T$ couplings to $f^{(\prime)}\anti f$
require a non-zero triplet vev. Thus, if the triplet vev is zero,
$H_T$ decays to fermion pairs will not be present at tree-level.

\item
There is a possibility of non-zero bi-lepton couplings of Higgs
bosons. For example, for the standard SU(2)$_L$ case,
with $Q=I_3+{Y\over 2}=-2$, the allowed 
doubly-charged cases are:
\begin{equation}
\begin{array}{l}
 e^-_Re^-_R\to \hmm(I=0,I_3=0,Y=-4) \,, \\ 
 e^-_Le^-_R\to \hmm(I={1\over 2},I_3=-{1\over 2},Y=-3) \,,\\  
 e^-_Le^-_L\to \hmm(I=1,I_3=-1,Y=-2)\,.
\end{array}
\label{helicitycases}
\end{equation}
Note that the above cases do not include the $I=3,Y=-4$ representation 
that yields $\rho=1$, nor the $I=1,Y=-4$ triplet with no neutral
member, but do include the $I=1/2,Y=-3$ doublet
representation with no neutral member, 
and the $I=1,Y=-2$ triplet representation. A $\nu_L\nu_L\to
\hzero(I=1,I_3=+1,Y=-2)$ coupling also exists, but does not lead
to neutrino mass if $v_L=\vev{\hzero}=0$ 
(as preferred for $\rho=1$ to be natural).

In the case of a $|Y|=2$ triplet representation
the lepton-number-violating coupling to (left-handed)
leptons is specified by the Lagrangian form:
\begin{equation}
{\cal L}_Y=ih_{ij}\psi^T_{iL} C\tau_2\Delta\psi_{jL}+{\rm h.c.}
\,,
\label{couplingdef}
\end{equation}
where $i,j=e,\mu,\tau$ are generation indices, 
the $\psi$'s are the two-component
left-handed lepton fields [$\psi_{\ell L}=\pmatrix{\nu_\ell,\ell^-}_L$], and
$\Delta$ is the $2\times 2$ matrix of Higgs fields:
\begin{equation}
\Delta=\pmatrix{\hm/\sqrt{2} & \hmm \cr \hzero & -\hm/\sqrt{2} \cr}\,.
\end{equation}

Limits on the $h_{ij}$ coupling strengths come from many sources. 
Experiments that
place limits on the $h_{ij}$ by virtue of the $\hmm\to
\ell^-\ell^-$ couplings include Bhabha scattering, $(g-2)_\mu$, 
muonium-antimuonium conversion, and $\mu^-\to e^- e^- e^+$.  
One finds rough limits of (for $\mhmm$ in GeV)
\begin{equation}
\begin{array}{ll}
|\hhmm_{ee}|^2\lsim 10^{-5} \mhmm^2 &
|\hhmm_{ee}\hhmm_{\mu\mu}|\lsim {\rm few}\times 10^{-5}\mhmm^2 \\
|\hhmm_{\mu\mu}|^2\lsim {\rm few}\times  10^{-5} \mhmm^2 &
|\hhmm_{e\mu}\hhmm_{ee}|\lsim {\rm few}\times  10^{-11}\mhmm^2 
\end{array}
\end{equation}
where $\hhmm_{ij}$ refers
to the $h_{ij}$ couplings as they appear in $\hmm$ interactions.
The last limit suggests
small off-diagonal couplings, as is assumed in what follows.
It is convenient to write
\begin{equation}
|\hhmm_{\ell\ell}|^2\equiv c_{\ell\ell} \mhmm^2(\gev)\,,
\label{hlimitform}
\end{equation}
where $c_{ee}\lsim 10^{-5}$ is the strongest of the limits.
The only constraint on the $h_{ij}$ through couplings
for the $\hzero$ and $\hm$ 
that is potentially stronger than those above 
goes away for $\langle\hzero\rangle= 0$, as required
in order to have $\rho=1$ naturally.

In left-right symmetric models, we must separate L from R and
use $Q=I_3(L)+I_3(R)+Y/2$, where $Y=B-L$, leading to
\begin{equation}
\begin{array}{l}
 e^-_Re^-_R\to \hmm(I(R)=1,I_3(R)=-1,Y=-2) \,, \\ 
 e^-_Le^-_R\to \hmm(I(R)=I(L)={1\over 2},I_3(R)=I_3(L)=-{1\over 2},Y=-2) \,,\\  
 e^-_Le^-_L\to \hmm(I(L)=1,I_3(L)=-1,Y=-2)\,.
\end{array}
\label{helicitycaseslr}
\end{equation}
respectively.
A $Y=-2$ bi-doublet, as required for the middle case, is not normally
considered, but is a logical possibility.
The $\call_Y$ for the 1st and 3rd cases above,
would be analogous to Eq.~(\ref{couplingdef})
with L fermions coupling to $\Delta_L$ and R fermions to $\Delta_R$.
In a large class of supersymmetric L-R symmetric models with
automatic R-parity conservation, the doubly-charged $\Delta_L$ triplet Higgs
bosons (and associated higgsinos) are naturally very light~\cite{mohapatra}.

\item
In the supersymmetric model context, the supersymmetric analogues
of the couplings listed above will all be present. 
In particular, any Higgs boson which couples to two fermions or two
vector bosons, will couple to their sfermion or gaugino partners.
Any Higgs boson which couples to another Higgs boson and a vector boson
will couple also to the higgsino and gaugino partners.
Any Higgs boson which couples to two other Higgs bosons will couple
to the corresponding higgsino partners. 
An $\hmm$ that couples to $\lmlm$ will couple to the corresponding
$\slep^-\slep^-$ channel. For example, the
$\hmm$ with $I(L)=1,I_3(L)=-1,Y=-2$ that is usually included
in a supersymmetric L-R symmetric model
will couple to $\slepl^-\slepl^-$.

If R-parity is violated in a supersymmetric model, still more couplings
of Higgs bosons emerge. The possible scenarios are quite complex.
In what follows, all supersymmetric couplings and consequent decays
of the Higgs bosons will be ignored.  Modifications to the phenomenological
discussions to come will, for the most part, be obvious.

\end{itemize}

It is now possible to outline the (non-supersymmetric)
decays of triplet Higgs bosons and their implications.
Obviously there is tremendous variation
according to which couplings are present. A particular focus
of the ensuing discussion will continue to 
be the exotic $\hmm$ that would constitute
an incontrovertible signature for an exotic Higgs representation
and which could be produced directly in $s$-channel $\lmlm$ collisions.
Regardless of how it is produced,
the decays of such a $\hmm$ would be very revealing.

\begin{itemize}

\item If the triplet vev is non-zero, then $H_T\to VV$ (possibly virtual)
decays and/or
$H_T\to f^{(\prime)} \anti f$ decays are usually most significant.

\item
If the triplet vev's are zero, then many channels are eliminated.
A type of mode that remains and is possibly important
is $H_T\to H_T^\prime V$. However, since many of
the $H_T$'s of a typical model are approximately degenerate, it is possible
for all such modes to be phase-space suppressed or virtual. 
In this case, bi-lepton decay modes
could dominate if the bi-lepton couplings are non-zero.
For example, in the case of a $I=1,|Y|=2$ $\hmm$, we have
\begin{equation}
\begin{array}{l}
\Gamma(\hmm\to\hm\wm)= {g^2\over 16\pi} {m_{\hmm}^3\beta^3
\over\mw^2}\sim (1.3\gev) \left({\mhmm\over
100\gev}\right)^3\beta^3\,, \\
\Gamma(\hmm\to\lmlm)=
{|\hhmm_{\ell\ell}|^2\over 8\pi} m_{\hmm}
\sim (0.4\gev) \left( {c_{\ell\ell}\over 10^{-5}}\right)
\left({\mhmm \over 100\gev}\right)^3\,.
\end{array}
\label{hmmwidths}
\end{equation}
where $\beta$ is the usual phase space suppression factor,
and Eq.~(\ref{hlimitform}) has been used.

The implications for the detection of the $\hmm$ have been considered
in \cite{trihtree}.
\begin{itemize}
\item
If one or more of the $c_{\ell\ell}$'s 
is significantly larger than $ 10^{-5}$, 
the $\ell^-\ell^-$ channel with the largest $c_{\ell\ell}$
will dominate $\hmm$ decays even if the $\hm\wm$ mode is allowed. 
Since there are currently no limits on $c_{\tau\tau}$, the $\tau^-\tau^-$
channel could easily have the largest partial width and be the dominant
decay of the $\hmm$.
\item
If the $\hm\wm$ mode is disallowed, then the $\ell^-\ell^-$ channel(s)
will provide the dominant signal
unless the $c_{\ell\ell}$'s are all extremely tiny.
\item
If {\it all} the $c_{\ell\ell}$'s are small and 
the $\hm\wm$ mode is allowed, then signatures would be more dilute
and more events would be required for detection.  
\item
If the $\hm\wm$ mode is virtual (\ie\ the $\hm$ and/or the $\wm$
must be off-shell) and the $c_{\ell\ell}$'s are all very small,
the $\hmm$ could be sufficiently long-lived to 
yield a vertexable track or escape the detector entirely.
(The latter requires that the $\hm\wm$ mode be doubly virtual
and that $c_{\ell\ell}\lsim 10^{-16}$.)
A search for stable particle highly ionizing tracks is then appropriate.
\end{itemize}

\item
For zero triplet vev's, the lightest $H_T$
will be stable unless it has bi-lepton couplings, in the absence of
which it must be neutral and would be a candidate for cold dark matter.

\end{itemize}

\section{Direct Production Probes}

Specific production processes of interest that are especially sensitive
to triplet Higgs representations can be separated
into those that do not require bi-lepton couplings and those that do.
Of course, the appropriate final state detection modes will depend upon
the relative weight of bi-lepton vs. other couplings.
In what follows, I once again focus on triplet representations.
The emphasis will be on $\lmlm$ and $\lmlp$ processes that can
take place at either an electron collider or a muon collider.

\bigskip
\noindent\underline{\bf Production processes independent of 
bi-lepton couplings.}
\bigskip

I give below an incomplete sampling of interesting processes that would probe
for the presence of an exotic Higgs representation.
\bit
\item 
Exclusive $\hmm\hpp$, $\hm\hp$, and $\hzero\hzero$
conjugate pair production in $\ell^+\ell^-$ or $pp$ collisions.
\eit

The cross section for pair production of a boson with 
weak isospin $I_3$ and charge
$Q$ and its conjugate via annihilation of a fermion with $i_3$ and $q$
and its anti-fermion partner is given by:
%\begin{eqnarray}
%\sigma^{\rm pair}&=\left( {\pi\alpha^2 \beta^3 s\over 6}\right)  &
%\Biggl\{ 2 Q^2q^2 P_{\gam\gam}+P_{\gam Z}{2Qq A(a_L+a_R)\over \xw\yw}
%\nonumber\\
%&&+P_{ZZ}{A^2(a_L^2+a_R^2)\over \xw^2\yw^2}\Biggr\}
%\,,
%\end{eqnarray}
\begin{equation}
\normalsize
\sigma^{\rm pair}=\left( {\pi\alpha^2 \beta^3 s\over 6}\right)  
\Biggl\{ 2 Q^2q^2 P_{\gam\gam}+P_{\gam Z}{2Qq A(a_L+a_R)\over \xw\yw}
+P_{ZZ}{A^2(a_L^2+a_R^2)\over \xw^2\yw^2}\Biggr\}
\,,
\label{pairsigma}
\end{equation}
where $\xw=\sin^2\thetaw$, $\yw=1-\xw$,
$A=I_3-\xw Q$, $a_L=i_3-\xw q$, $a_R=-\xw q$, $P_{\gam\gam}=s^{-2}$,
$P_{ZZ}=[(s-\mz^2)^2+\mz^2\gamz^2]^{-1}$, and 
$P_{\gam Z}=(s-\mz^2)P_{ZZ}/s$. A color averaging factor of 1/3
must be supplied in the context of $pp$ or $p\anti p$ collisions
when convoluting this pair cross section with quark distribution functions;
a factor of 1/2 is to be included for $\hzero\hzero$ production.
Note that the pair cross sections do not rely on there being a non-zero
vev for a neutral Higgs field.

\begin{figure}[ht]
\leavevmode
\begin{center}
%%\resizebox{!}{11cm}{%
%%\includegraphics{layout.eps}}
\epsfxsize=3.5in
\hspace*{0.25in}
\epsffile{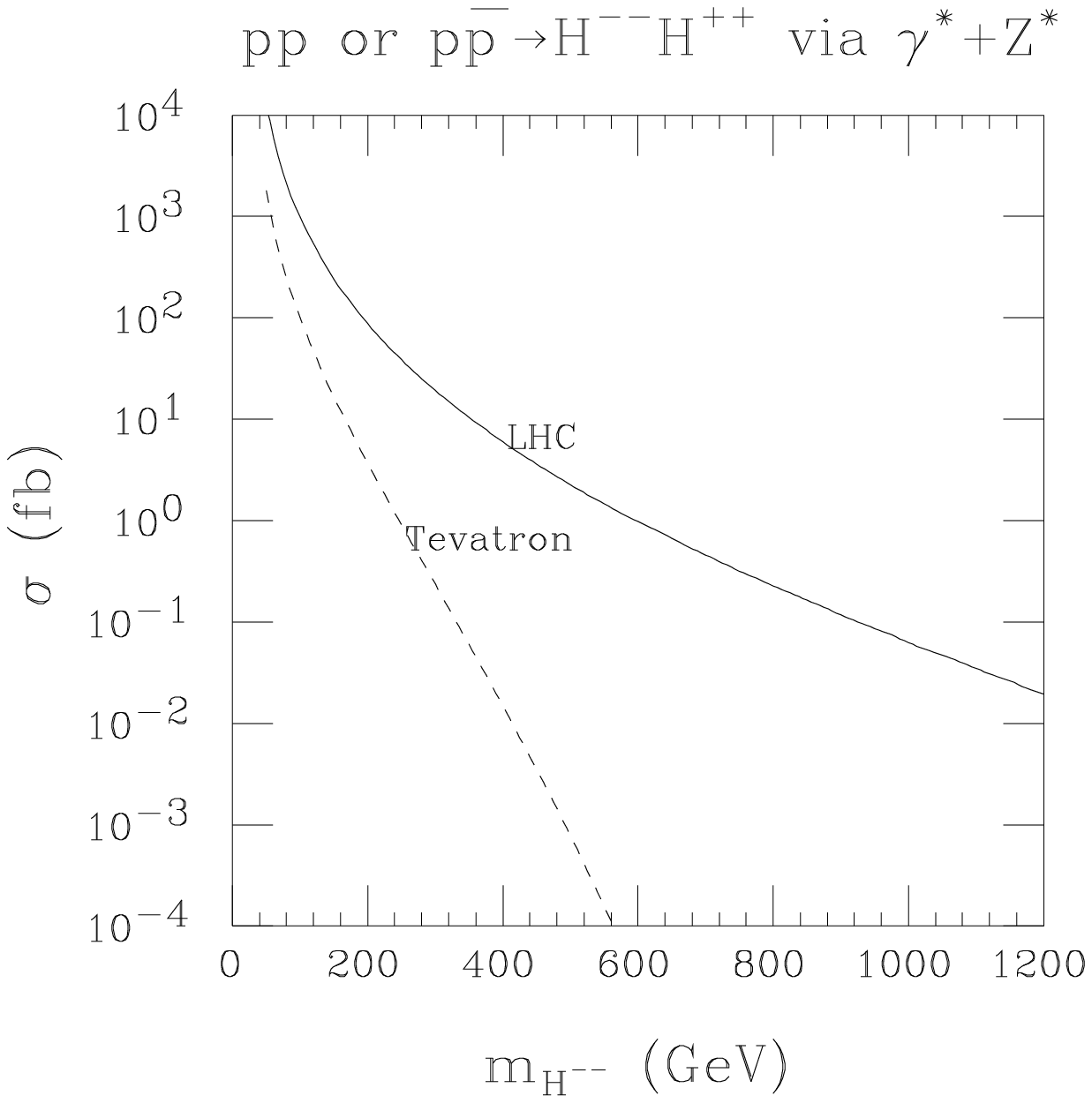}
\end{center}
\fcaption{$\hpp\hmm$ pair production
cross section as a function of $\hmm$ mass
for the Tevatron and LHC.}
\label{fig:xsec}
\end{figure}

Signatures of the $\hmm\hpp$ final state can be extremely striking.
If $\hmm\to\ell^-\ell^-$ decays are dominant, 
for $\ell=e$ or $\mu$ only a few
spectacular events with two like-sign lepton pairs
of equal mass are needed for an unambiguous signal. 
If $c_{\tau\tau}$ is the largest bi-lepton coupling,
the $4\tau$ final state can be isolated using relatively simple cuts.
The detection limits for $\hmm\hpp$ pair production in the case
of a T$_2$ representation have been explored \cite{kpitts}. The
Tevatron and LHC production cross sections are given in Fig.~\ref{fig:xsec}.
When the $\hmm$ decays are dominated by $\ell^-\ell^-$ decay modes,
Ref.~\cite{kpitts} finds that detection of $\hmm\hpp$ pair production
at the Tevatron
(operating at $\rts=2\tev$ with $L=30\fbi$) will be possible for $\mhmm$ 
up to $300\gev$ for  $\ell=e$ or $\mu$ and  $180\gev$ for $\ell=\tau$.
The corresponding limits at the LHC are estimated by requiring
the same raw number of events before cuts and efficiencies as
needed at the Tevatron --- $\sim 10$ for $\ell=e,\mu$ and $\sim 300$ for
$\ell=\tau$ --- yielding $\mhmm$ discovery up to roughly $925\gev$ ($1.1\tev$)
for $\ell=e,\mu$ and $475\gev$ ($600\gev$) for $\ell=\tau$,
assuming total integrated luminosity of $L=100\fbi$ ($L=300\fbi$).
For $\ell=e,\mu$, the reach of the 
LHC detectors will likely be even greater than
this, due to the improved lepton acceptance and resolution
anticipated over the current generation of hadron collider detectors.
For $\ell=\tau$, this simple extrapolation may not account
for a different signal-to-background ratio in 
$\tau$ selection at
the LHC.  A more complete study is necessary to evaluate this.
If the $\hmm\to \hm\wm$ decay is dominant, then the signals will be more
dilute than for the $e^-e^-$ or $\mu^-\mu^-$ cases. Using the leptonic
($\nu\ell$, $\ell=e,\mu$)
decays of the $\wm\wp$ in the $\hmm\hpp\to \hm\wm\hp\wp$ final state
suggests that one can
hope to achieve $S/\sqrt B$ levels similar to the $\tau^-\tau^-$ case,
in which case the discovery reaches for the latter channel
discussed above would apply.

Although observation of $\hmm\hpp$ pair production 
is certainly the clearest signal for an exotic Higgs representation,
additional information and confirmation can be obtained from
the $\hp\hm$ and $\hzero\hzero$ production rates,
which will reveal non-doublet correlations
between $Q$ and $A=I_3-\xw Q$. Of course,
in these latter cases, decays could be more complicated and would need
to be understood in order to extract the underlying pair cross section.

In any case, the conclusion is that if there are exotic Higgs triplet
bosons, they will be observed
in pair production at the LHC, if not at the Tevatron, up to the highest
mass ($\sim 500\gev$) for which they could be directly produced
in the $s$-channel at a first $\lmlm$ collider with $\rts\lsim 0.5\tev$.
\bit
\item Exclusive production of non-conjugate Higgs pair.
\eit

Non-conjugate Higgs pairs can be produced via $s$-channel virtual
$Z$ exchange in $\epem$ collisions and via virtual $Z$ or $W$ exchange
in $pp$ collisions. However, there has not been a careful study of the signals
and backgrounds to the many possible final state detection channels.
One background free channel would be $\wm\to \hmm\hp$ production
in which the $\hmm$ decays to $\ell^-\ell^-$ ($\ell=e,\mu,\tau$).  
If the $\hp$ also belongs to the triplet representation,
the required coupling is sizeable and independent of the triplet vev. This
mode will generally allow confirmation of any signal seen
for $\hmm\hpp$ pair production at the Tevatron or LHC (see above).
\bit
\item Exclusive $\hp\wm$, $\hm\wp$ production.
\eit

The only diagram is $s$-channel $Z$ exchange.
A non-zero $Z\to \hpm\wmp$ coupling requires a (L) {\bf triplet}
representation with a non-zero vev. Signal and background rates
for this reaction at an $\epem$ collider have 
been explored in Refs.~\cite{cheungetal,godboleetal}.
\bit
\item $\wm\wp$ production from $\wm\wp$ fusion in $\lmlp$ collisions.
\eit

The influence of triplet Higgs representations on
standard $\wm\wp\to\wm\wp$ fusion was considered at some length
in \cite{trihtree}, as summarized in \cite{hhg}. In general,
the $WW$ scattering processes, of which this is our first example,
will exhibit strong unitary-violating behavior unless an appropriate
set of Higgs bosons are present. Any Higgs boson belonging
to a representation in which there is a non-zero vev for a neutral
member will generally contribute to one or more of the amplitudes
for $WW$ scattering. 
Contributions to the $\wm\wp\to\wm\wp$ amplitude that do
not arise from standard electroweak interactions include:
\begin{enumerate}
\item
$s$-channel $\hzero$ exchanges (appropriate vev required).
\item
$t$-channel $\hzero$ exchanges (appropriate vev required).
\item
$u$-channel $\hpp$ exchanges (vev and {\bf triplet} required).
\end{enumerate}
Note that only the $u$-channel diagram requires a triplet field; 
contributions to the
$s$ and $t$ channel amplitudes are possible for both doublet and
triplet neutral Higgs bosons, provided the appropriate 
doublet and triplet (respectively)
vev's are non-zero.
In a general model, multiple exchanges of each type are possible.
All Higgs bosons contributing to the amplitude must have mass 
$\lsim 1\tev$ in order to avoid non-unitary high energy behavior.
For lepton colliders, the draw-back of this scattering process 
and the other reactions summarized
below, is that $\rts>1\tev$ is needed to fully 
explore the nature of the contributing diagram amplitudes.
\bit
\item $\wm\wm$ production from $\wm\wm$ fusion in $\lmlm$ collisions.
\eit

This process was considered, for example, in Refs.~\cite{trihtree,hanemem}.
Contributions to the $\wm\wm\to\wm\wm$ amplitude are: 
\begin{enumerate}
\item
$s$-channel $\hmm$ exchanges (vev and {\bf triplet} required).
\item
$t$-channel $\hzero$ exchanges (appropriate vev required).
\item
$u$-channel $\hzero$ exchanges (appropriate vev required).
\end{enumerate}
The same comments as above apply. In particular, non-zero contributions
require an appropriate non-zero vev for a neutral Higgs field.

Let us give a specific example in this case. 
In the SM, there are two $\hsm$-exchange graphs: one is a $t$-channel
and the other a $u$-channel graph.  They can be thought of as combining
together and having effective strength $g^2\mw^2$, as required to
cancel the bad high energy behavior coming from graphs involving standard
electroweak boson exchanges. Now
consider the 1D$_1$+1T$_2$+1T$_0$(real) Higgs 
sector detailed in the Introduction. In this model, there are
three $u$-channel and three $t$-channel graphs for the 
$H_1^0$, $H_1^{0\,\prime}$
and $H_5^0$ neutral Higgs bosons and an $s$-channel
graph for the $H_5^{--}$. An $s$-channel graph is equivalent to the sum of a
$t$- and a $u$-channel graph except for an overall sign. Thus, the
effective contributions of the different Higgs exchanges are:
\bea
&H_1^0:~~g^2c_H^2\mw^2\,; & H_1^{0\,\prime}:~~(8/3)g^2s_H^2\mw^2\,; \nonumber\\
&H_5^0:~~(1/3)g^2s_H^2\mw^2\,; & H_5^{--}:~~-2g^2s_H^2\mw^2\,, 
\label{wmwmcontributions}
\eea
where $s_H\equiv\sin\theta_H$ ($\theta_H$ was defined earlier). 
Note that these always
sum to the SM weight of $g^2\mw^2$, thereby guaranteeing good high
energy behavior.  Also note that in the limit of $s_H\to 0$ (vanishing
vev's for the neutral triplet members) only the $H_1^0$ amplitudes
survive; indeed, in this limit the $H_1^0$ has the same properties
as the SM $\hsm$.  However, for substantial $s_H$ the triplet Higgs
fields play a very important role in guaranteeing good high-energy
behavior for the $\wm\wm\to\wm\wm$ amplitude. Further, for sizeable
$s_H$, the $H_5^{--}$ will
create a clear resonance bump in the $\wm\wm$ invariant mass spectrum
that should be easily detectable at an $\lmlm$ accelerator 
of sufficient energy (and also in $\wm\wm$ fusion at the LHC).
\bit
\item $\hm\hp$ from $\wm\wp$ fusion in $\lmlp$ collisions.
\eit

This is our first example of a process 
involving a mixture of Higgs bosons and $W$ bosons.
For such processes, not all amplitude contributions require 
that the neutral member of the Higgs boson representation have a non-zero vev. 
Thus, in what follows the exchanges which do and don't require
such a vev are indicated. 
For $\wm\wp\to\hm\hp$, the contributing diagrams include:
\begin{enumerate}  
\item
$s$-channel $\hzero$ exchanges (appropriate vev required);
\item
$t$-channel $\hzero$ and $\a^0$ exchanges (vev not required);
\item
$t$-channel $Z$ exchange (vev and {\bf triplet} required);
\item
$u$-channel $\hmm$ exchange ({\bf triplet}, but not vev, required);
\item
quartic coupling.
\end{enumerate}
The only standard electroweak contribution to $\wm\wp\to\hm\hp$
is $s$-channel $Z$-exchange.
\bit
\item $\hm\hm$ from $\wm\wm$ fusion in $\lmlm$ collisions.
\eit
The diagrams that contribute to $\wm\wm\to\hm\hm$ include:
\begin{enumerate}
\item
$s$-channel $\hmm$ exchange (vev and {\bf triplet} required);
\item
$t$-channel $\h^0$ and $\a^0$ exchanges (vev not required); 
\item
$t$-channel $Z$ exchange (vev and {\bf triplet} required);
\item
$u$-channel $\h^0$ and $\a^0$ exchanges (vev not required); 
\item
$u$-channel $Z$ exchange (vev and {\bf triplet} required);
\item
quartic coupling.
\end{enumerate}
This reaction has been especially thoroughly studied in Ref.~\cite{rizzoemem},
where substantial sensitivity to the exact nature of the Higgs sector
was demonstrated.
\bit
\item $\hp\wm$ or $\wp\hm$ production via $\wp\wm$ fusion in $\lplm$ collisions.
\eit

Contributing diagrams include: 
\begin{enumerate}
\item
$s$-channel $Z$ exchange (vev and {\bf triplet} required);
\item
$t$-channel $\hzero$ exchange (vev required);
\item
$u$-channel $\hpp$ exchange (vev and {\bf triplet} required);
\end{enumerate}

Even this partial list (which, for example, does
not include processes with $Z$'s in the final state)
should make clear that there are a very large
number of processes to be considered.  Because of the lurking
possibility of unitarity-violating high-energy behavior for
the processes considered above, the exact energy dependence
and cross section magnitudes are quite sensitive to the 
contributing Higgs bosons through their masses and couplings.
Thus, a full study of {\it all}
such processes has an excellent chance of leading
to a full understanding of a complicated
Higgs sector.  In a lepton collider environment, it will be absolutely crucial
to have both $\ell^-\ell^-$ and $\ell^-\ell^+$ collisions
in order to be able to study all of the reactions of interest.

\bigskip
\noindent\underline{\bf Production processes requiring bi-lepton couplings.}
\bit
\item Single ($s$-channel) $\hmm$ production in $\lmlm$ collisions
\cite{jfgemem,frampton,cuypers,huituemem,alanakyan}.
\eit

Let us assume that the $\hmm$ has already been discovered,
for example at the LHC via $\hmm\hpp$ pair production as described earlier.
If the $\hm$ is detected in the $\lmlm$ final state, study of the $\hmm$
via $\lmlm$ collisions will have an extremely high priority.
If the error in $\mhmm$ as measured at the LHC, $\Delta\mhmm$,
is sufficiently small, a narrow scan 
at the $\lmlm$ collider will, in most cases, allow us to center on 
and study the resonance in a factory-like setting even if
the coupling, $c_{\ell\ell}$,  responsible for the production is very small.

The crucial feature of the $\lmlm$ collider for narrow resonance production
in this context is the width of and 
luminosity contained in the narrow Gaussian-like peak\footnote{In addition 
to the narrow Gaussian-like peak, there is a long tail at $\sqrt s$
values below the nominal central value coming from beamstrahlung 
and bremsstrahlung. When we refer to the luminosity in the Gaussian-like peak,
this does not include the luminosity contained in this tail. The losses
to this tail are larger for electron colliders than muon colliders.}~ 
centered at the nominal $\rts$ setting of the machine. We will denote
the rms width of this Gaussian-like peak by $\sigrts$. It is related
to the intrinsic (\ie\ before including
beamstrahlung and bremsstrahlung) machine beam energy resolution $R$ by
\begin{equation}
\sigrts\sim 0.2\gev \left({\mhmm\over 100\gev}\right)
\left({R\over 0.2\%}\right)\,,
\label{sigmamagnitude}
\end{equation}
Values of $R$ that can be achieved at an electron collider are
$R\sim 0.2\%-0.3\%$; $R\sim0.1\%$ is easily achieved at a muon collider,
with values as small as $R\sim 0.003\%$ possible at reduced luminosity.
As described below, the cross section for $\hmm$ production is
maximized if $\sigrts\lsim\gamhmm$ can be achieved.

The effective cross section, denoted by $\sighmmbar$, for $\lmlm\to\hmm$
is obtained by convoluting the standard $s$-channel resonance form
with the distribution in $\sqrt s$ of the machine luminosity. 
Results for the Gaussian approximation are easily summarized
in the two limits of $\gamhmm \gg \sigrts$ and $\gamhmm\ll\sigrts$. 
Taking $\sqrt s=\mhmm$,  $\sighmmbar$ is given by:
\begin{equation}
\sighmmbar =\left\{ 
\begin{array}{ll} 
{4\pi \br(\hmm\to \emem)\over \mhmm^2}\,,
&\mbox{$\gamhmm\gg\sigrts$;} \\
{\sqrt{\pi}\over 2\sqrt{2}}\ 
{4\pi {\Gamma(\hmm\to \emem)\over\sigrts}\over \mhmm^2}, 
&\mbox{$\gamhmm\ll\sigrts$\, .}
\end{array}
\right.
\label{hmmxsec}
\end{equation}
We compute rates as $L\sighmmbar$ where $L$ is the luminosity
within the Gaussian peak (after accounting for losses to the radiative tail).
Two-year integrated luminosity of $L=50\fbi$ is expected at an $\emem$ collider.
At a $\mummum$ collider, the expected $L$ depends strongly on both $R$ and
$\rts$. Conservative two-year expectations are $L=0.1,0.2,1.2\fbi$ 
for $R=(0.003,0.01,0.1)\%$ at $\rts\sim 100\gev$ rising to
$L=(2,6,14)\fbi$ at $\rts\sim (200,350,400)\gev$, assuming $R\sim 0.1\%$.
As already noted, initially it will be necessary to scan
over the interval $\mhmm\pm\Delta\mhmm$ in order to center on
$\rts\simeq\mhmm$, implying that the resonance peak will
receive only a fraction of the total $L$.

In fact, designing an appropriate scan is not necessarily straightforward.
Even if the $\hmm$ is observed at the LHC in the $\lmlm$
final state, the manner in which to perform a narrow scan at an $\lmlm$
collider for centering on $\rts\simeq\mhmm$ can be delineated only
if $\gamhmm$ can be measured or estimated from the LHC observations.
This is because the luminosity per scan point needed to observe
or exclude the $\hmm$ depends on $\gamhmm$, see Eq.~(\ref{hmmxsec}).
The most optimistic case is that the width of the $\hmm$
is larger than the intrinsic resolution (expected to be of order a
few percent of $\mhmm$) for 
the $\emem$ and/or $\mummum$ final state in which it is observed at the LHC.
Then, $\gamhmm$ can be read directly off the
final state mass distribution(s).
Another possibility is that the $\hmm\to\wm\hm$ decay is observed.
In this case, even if the mass resolution in all
final states is inadequate for a direct $\gamhmm$ determination,
we can compute the $\wm\hm$ partial width (using the presumably
measured $\hm$ mass) up to a Higgs-representation-dependent factor.  Then
an estimate for the total width is given by $\gamhmm\sim
\Gamma(\hmm\to\hm\wm)/\br(\hmm\to\hm\wm)$.
In either case, $\br(\hmm\to\lmlm)$ would have been measured
at the LHC and we could compute 
$\Gamma(\hmm\to\lmlm)=\gamhmm\br(\hmm\to\lmlm)$; these
are the crucial ingredients in Eq.~(\ref{hmmxsec}). 
If $\gamhmm\gg\sigrts$,
we would place scan points within $\pm \Delta\mhmm$ of $\mhmm$
at intervals of roughly $\gamhmm$ and devote
the appropriate luminosity per point, as determined by $\br(\hmm\to\lmlm)$,
to ensure that a signal is seen if present at that scan point.
If $\gamhmm\ll\sigrts$, we would place points at intervals of roughly $\sigrts$
and the required luminosity per point would be determined by 
$\Gamma(\hmm\to\lmlm)/\sigrts$.

If, at the LHC, the $\hmm\to\hm\wm$
decay is not observed and the bi-lepton mass distributions are consistent
with resolution, it will not be possible to ab initio set up a 
scan procedure that is guaranteed to find the $\hmm$. The problem is that 
$\gamhmm$ could be $\ll\sigrts$, in which case knowledge of
$\Gamma(\hmm\to\lmlm)$ is required [see Eq.~(\ref{hmmxsec})] 
to determine the luminosity per scan
point needed to exclude or discover the $\hmm$. This situation applies
also if the representation-dependent value for $\Gamma(\hmm\to\hm\wm)$ 
computed (using an LHC measurement of the $\hm$ mass) is $\ll\sigrts$.
This is because $\Gamma(\hmm\to\lmlm)$ cannot be estimated from
the computed $\Gamma(\hmm\to\hm\wm)$ value, there being no LHC measurement of
$\br(\hmm\to\hm\wm)/\br(\hmm\to\lmlm)$. In both the above cases,
the best that one could do is to spread a certain amount of
luminosity (say one year's worth) over the $\mhmm\pm\Delta\mhmm$ mass interval
at intervals of $\sigrts$. As will become apparent later, this 
would still allow $\hmm$ detection in $\lmlm$
collisions for quite small $c_{\ell\ell}$ values.

Although the need to scan in order to center
on $\rts\simeq\mhmm$ is an important ingredient in assessing our ability to
study the $\hmm$ in $\lmlm$ collisions, a full discussion 
of the various scan procedures is quite involved.  For simplicity,
I discuss here only the event rates that would arise for a scan
point placed successfully at $\rts\simeq\mhmm$. It is worth
re-emphasizing that the $L$ per scan point might be a possibly
small fraction of the full $L$ available. 
The discussion divides
naturally into the $\gamhmm\gg\sigrts$ and $\gamhmm\ll\sigrts$ cases.

\begin{description}
\item{{\bf A:}} $\gamhmm\gg\sigrts$. 
\end{description}

This scenario is typical if the $\hmm\to\wm\hm$ decay is allowed or if one or
more of the $c_{\ell\ell}$'s is above $10^{-5}$. 
For $\rts\simeq\mhmm$, the number of $\hmm$ events in $\lmlm$ collisions is
\begin{equation}
N(\hmm)\sim 2.5\times 10^{9} \left({100\gev\over\mhmm}\right)^2
\left({L\over 5\fbi}\right) \br(\hmm\to \lmlm)\,.
\label{eventratebroad}
\end{equation}
If the $\hmm\to\hm\wm$
decay mode dominates the total width, then $\br(\hmm\to\lmlm)\sim 0.3
\beta_{\wm\hm}^{-3}(c_{\ell\ell}/10^{-5})$.  For 
$\beta_{\hm\wm}=1$, $c_{\ell\ell}\sim 10^{-5}$  and $\mhmm=500\gev$, 
we predict $3\times 10^7$ $\hmm$ events (dominated by the $\hm\wm$ final state)
for a typical scan point luminosity of $L=5\fbi$. 
(This large an integrated luminosity per scan point is more easily
achieved at an $\emem$ collider than at a $\mummum$ collider, even though
the latter would 
be operated with $R=0.1\%$ when $\gamhmm$ is large.)

As $\beta_{\hm\wm}$ decreases below 1, the number of events grows
rapidly.\footnote{This growth occurs until $\gamhmm$ approaches
$\sigrts$.}~  
A total of 100 $\hmm$ events are produced for
\begin{equation}
\left.c_{\ell\ell}\right|_{\rm 100~events}
=1.3\times 10^{-12} \left({\mhmm\over100\gev}\right)^2\beta_{\hm\wm}^3
\left({5\fbi\over L}\right)
\,,\quad \gamhmm\gg\sigrts\,;
\end{equation}
with $L=5\fbi$ per scan point
an observable signal occurs for very small 
$c_{\ell\ell}$ values.

If the $\tau\tau$ decay mode dominates $\gamhmm$, then $\br(\hmm\to\lmlm)\sim
c_{\ell\ell}/c_{\tau\tau}$. For $L=5\fbi$, $10^7$ $\hmm$ events
(almost entirely $\tau^-\tau^-$) would be 
obtained for $c_{\ell\ell}/c_{\tau\tau}=0.1$
and $\mhmm=500\gev$. In this case, 100 $\hmm$ events
would correspond to $c_{\ell\ell}/c_{\tau\tau}=4\times 10^{-8} 
(\mhmm/100\gev)^2$,
again a very respectable sensitivity. Note that the phenomenology of this 
latter case of $\tau^-\tau^-$ dominance of $\hmm$ decays 
is essentially independent of $\Gamma(\hmm\to\hm\wm)$.

Extracting the actual magnitudes of the $c_{\ell\ell}$ coupling
strengths from the $\hmm$ observations would
be a very important goal.  In assessing our ability to
perform this task, we shall presume that the efficiencies for the
observed final states after cuts can be computed with sufficient reliability
that the total number of $\hmm$ events before cuts (summing over all modes)
and the relative branching fractions $\br(\hmm\to X)$ for the different
observed final states $X$ can all be determined with reasonable accuracy. 
The following procedures
would then be appropriate in the present $\gamhmm\gg\sigrts$ case.
\begin{description}
\item{(a)} 
The partial width for any final state $X$ can be computed from $\gamhmm$
(which would have been directly measured as part of the scan in this case,
if not already known from LHC measurements)
as $\Gamma(\hmm\to X)=\gamhmm\br(\hmm\to X)$. If 
$X=\hm\wm$ is observed, then a Higgs-representation-dependent value
of $\Gamma(\hmm\to\hm\wm)$ can be computed and compared to the direct
determination.
\item{(b)}
The value of $\br(\hmm\to\lmlm)$ determined from $N(\hmm)$ 
[see Eq.~(\ref{eventratebroad})] can be used to compute
$\Gamma(\hmm\to\lmlm)=\br(\hmm\to\lmlm)\gamhmm$ and, 
thence, $c_{\ell\ell}$, even if the $\lmlm$
final state cannot be detected.
\item{(c)}
If the $\lmlm$ decay mode can be seen but is not
dominant, then the above $\Gamma(\hmm\to\lmlm)=\br(\hmm\to\lmlm)\gamhmm$ 
determination can be used to compute
$\Gamma(\hmm\to X)=\Gamma(\hmm\to\lmlm)\br(\hmm\to X)/\br(\hmm\to\lmlm)$
for the other observed $\hmm$ final states $X$ as a cross check
of the determinations in (a). 
\end{description}

\begin{description}
\item{{\bf B:}} $\gamhmm\ll\sigrts$. 
\end{description}

If $\hmm\to\hm\wm$ is either highly suppressed or virtual, then $\gamhmm$
can be very small if all of the $c_{\ell\ell}$'s
are relatively small. At a $\mummum$ collider, $\sigrts = 3
\mev\times (\mhmm/100\gev)$ can be achieved for $R=0.003\%$, which implies [see
Eq.~(\ref{hmmwidths})] that all $c_{\ell\ell}$'s would have to be below
$10^{-7}\times (\mhmm/100\gev)^3$ before this situation would be forced upon us.

Using Eq.~(\ref{sigmamagnitude}) for
$\sigrts$ and the earlier result for $\Gamma(\hmm\to \lmlm)$,
Eq.~(\ref{hmmxsec}) predicts an event rate of
\begin{equation}
N(\hmm)\sim 3\times 10^{9}\left({c_{\ell\ell}\over 10^{-5}}\right)
\left({0.2\%\over R}\right)\left({L\over 5\fbi}\right)\,.
\label{eventratenarrow}
\end{equation}                
Thus, an enormous event rate results if $c_{\ell\ell}$ is within a few
orders of magnitude of its upper bound.
In this $\gamhmm\ll\sigrts$ case, 
the total number of events depends only on $c_{\ell\ell}$.
From Eq.~(\ref{eventratenarrow}), we predict 100 $\hmm$ events
for
\begin{equation}
\left.c_{\ell\ell}\right|_{\rm 100~events}\sim 3.3\times 10^{-13} 
\left({R\over 0.2\%}\right)\left({5\fbi\over L}\right)
\,,\quad \gamhmm\ll\sigrts\,.
\label{narrow100}
\end{equation} 
How many events are needed for observation will depend upon the final state.

What final states will be important? Assuming that $\hm\wm$ is two-body
forbidden [as typically required for small $\Gamma(\hmm\to\hm\wm)$,
see Eq.~(\ref{hmmwidths})], 
one or several of the $\ell^-\ell^-$ modes would dominate
unless all the $c_{\ell\ell}$'s are extremely small, in which case
the ${\hm}^*\wm$, $\hm{\wm}^*$ semi-virtual three-body modes would be dominant.
The precise cross-over point between the $\ell^-\ell^-$ modes and the
semi-virtual modes depends on details.
Of course, if the $\hmm$ is observed at the LHC or NLC, we will know ahead
of time what final state to look in and its detailed
characteristics, even if the semi-virtual final state is dominant.
Only the latter semi-virtual modes and the $\lmlm$ final state
would have significant backgrounds at an $\lmlm$ collider.  

How many events are required for observation of the $\hmm$? The semi-virtual
$\hm\wm$ final state would probably have the least distinctive
signature, and
we could pessimistically assume that 1000 events would be required for
$\hmm$ detection if this is the dominant final state mode. The $\lmlm$
final state (same $\ell$ as for the initial state) would also have a large
background.  Presumably at least 100 events would be needed before applying
cuts to tame the direct $\lmlm\to\lmlm$ background. Perhaps a similar number
would be required for the $\hm\wm$ final state when it is not virtual
and for the $\tau^-\tau^-$ final state. Fewer than 10 events would presumably
suffice for the $\mummum$ final state in $\emem$ collisions and the $\emem$
final state in $\mummum$ collisions. If one adopts the prejudice that
$c_{ee}\ll c_{\mu\mu}$ and $c_{\tau\tau}$, then the $\mummum$ and
$\tau^-\tau^-$ final states would probably dominate $\hmm$ decays
and 10 to 100 events would suffice in $\emem$ collisions.

Even if 1000 events are required, 
Eq.~(\ref{narrow100}) implies dramatic sensitivity.  
For $L=5\fbi$ per scan point and requiring 1000 events, in $\emem$ collisions
we are able to observe a signal for $c_{ee}$ values that are
nearly 7 orders of magnitude smaller than current limits.
We can probably do even
better in the case of $c_{\mu\mu}$ using $\mummum$
collisions, depending upon just how low in $R$ one can go without losing
too much luminosity in this latter case.

Determination of the basic partial widths of the $\hmm$
follows a slightly different strategy in this $\gamhmm\ll\sigrts$ case 
as compared to the $\gamhmm\gg\sigrts$ case.
As before we presume that the total event rate and the branching ratios
for observable final states can be obtained with reasonable accuracies.
\begin{description}
\item{(a)} 
When $\gamhmm\ll\sigrts$, Eq.~(\ref{eventratenarrow}) shows that the total
event rate is a direct measure of $c_{\ell\ell}$ [equivalently
$\Gamma(\hmm\to\lmlm)$].
\item{(b)} If the $\lmlm$ final state is observable, we can then
compute $\Gamma(\hmm\to X)=\Gamma(\hmm\to\lmlm)\br(\hmm\to
X)/\br(\hmm\to \lmlm)$ for other observable channels $X$.
If $X=\mupmum$, $\tauptaum$ is observed, 
this yields $c_{\mu\mu}$, $c_{\tau\tau}$. If $X=\hm\wm$ is observed,
the value of $\Gamma(\hmm\to\hm\wm)$ can be compared
to predictions for different Higgs representation choices. Finally,
$\gamhmm$ can be computed by summing all the partial widths determined
as above. If $\gamhmm$ is not a great deal smaller than $\sigrts$,
a very precise scan of the $\hmm$ resonance
might allow a direct measurement of $\gamhmm$
which could then be compared to the value computed above.
\item{(c)} If the $\lmlm$ final state is not detected, and
$\gamhmm$ is so much smaller than $\sigrts$ that any sort of
scan determination of $\gamhmm$ is impossible, then a
direct determination
of the partial widths for $X\neq\lmlm$ channels will not be possible.
In this case, the best that one can hope for is that the
$\hm\wm$ final state is detected, in which case the $\Gamma(\hmm\to\hm\wm)$
partial width can be computed using an assumed representation for the $\hmm$
and other partial widths can be obtained from the ratio of branching ratios.
\end{description}

Aside from direct $s$-channel resonance production, there are
several other potentially interesting processes.
\bit
\item Exclusive $\hmm Z$ and $\hmm \gam$ production in $\lmlm$ collisions.
\eit

The reaction $\lmlm\to\hmm\gam$ 
is essentially equivalent to using the bremsstrahlung tail 
in $\rts$ at the $\lmlm$ collider to self-scan for the $\hmm$.
The $\hmm\gam$ rate is proportional to $c_{\ell\ell}$
as in the $\gamhmm\ll\sigrts$ limit of on-resonance production,
but observable rates are only possible if $c_{\ell\ell}$ is not too far below
current bounds. If the $\hmm$ is discovered in this way,
first measurements of $\mhmm$ and $c_{\ell\ell}$ would be
obtained and one could then proceed to the $s$-channel scan for centering 
on $\rts\simeq\mhmm$ as described above and a factory-like study
of the $\hmm$.

\bit
\item Exclusive $\hm\wm$ production in $\lmlm$ collisions.
\eit

This process can occur via a diagram with $\nu_{\ell}$
exchanged in the $t$-channel, provided there is
a $\hm\to \ell^-\nu_{\ell}$ coupling. If the $\hm$ 
is part of a triplet representation, such a coupling would be automatically
present in Eq.~(\ref{couplingdef}). The $\lmlm\to \hm\wm$ production
would be both an interesting way of observing any $\hm$ with 
a sizeable value for such a coupling
and a way of determining the magnitude of the coupling.

\bit
\item The value of polarization.
\eit

For all the above triplet Higgs production mechanisms 
mediated by a bi-lepton coupling,
it would be very valuable to have the ability
to polarize one or both of the lepton beams. The polarization
dependence of the rates can allow a model-independent determination
of the representation to which the triplet Higgs boson belongs.
In particular, Eqs.~(\ref{helicitycases}) and (\ref{helicitycaseslr}) show 
the specific polarization choices for the incoming $\ell^-$ beams that yield
a non-zero rate for $\lmlm\to \hmm$ $s$-channel production 
for different choices of the $\hmm$ representation.
For example, we see that the production rate for
a $Y=-2$ triplet $\hmm$ would be suppressed if either of the $\ell^-$ beams
is given right-handed polarization, since such an $\hmm$
couples only to $\ell^-_L\ell^-_L$.

\section{\bf Conclusions}

The Higgs sector remains one of the big unknowns of particle physics.
Although, it could consist of only doublets and singlets,
more exotic Higgs sectors remain a distinct possibility and are even
required in some popular extensions of the Standard Model,
in particular left-right symmetric models. In this review, 
we have delineated the essential role that an
$\lmlm$ collider would play in
fully revealing the nature of a typical exotic Higgs sector,
focusing on the case in which Higgs triplet fields are present.
It would appear that construction of appropriate
$\lmlm$ accelerators (both $\ell=e$ and $\ell=\mu$) could
become a very high priority in order to ensure our ability to
fully explore the exact nature of an exotic Higgs 
sector that is first revealed by LEPII, Tevatron or LHC experiments.

\nonumsection{References}

\end{document}